\documentclass[reprint,twocolumn,amsmath,amssymb,aps,superscriptaddress]{revtex4-1}
\usepackage{graphicx}
\usepackage{dcolumn}
\usepackage{bm}
\usepackage{float}
\usepackage{amssymb}
\usepackage{color}
\usepackage{multirow}
\usepackage{stmaryrd}
\usepackage{ulem}
\begin{document}


\title{Enhanced Anomalous Hall Effect in Magnetic Topological Semimetal $\mathbf{Co}_3\mathbf{Sn}_{2-x}\mathbf{In}_{x}\mathbf{S}_2$}

\author{Huibin Zhou}
\affiliation{International Center for Quantum Materials, School of Physics, Peking University, China}

\author{Guoqing Chang}
\affiliation{Laboratory for Topological Quantum Matter and Advanced Spectroscopy(B7), Department of Physics, Princeton University, Princeton, NJ 08544, USA}

\author{Guangqiang Wang}
\affiliation{International Center for Quantum Materials, School of Physics, Peking University, China}

\author{Xin Gui}
\affiliation{Department of Chemistry, Louisiana State University, Baton Rouge, LA 70803, USA}

\author{Xitong Xu}
\affiliation{International Center for Quantum Materials, School of Physics, Peking University, China}

\author{Jia-Xin Yin}
\affiliation{Laboratory for Topological Quantum Matter and Advanced Spectroscopy(B7), Department of Physics, Princeton University, Princeton, NJ 08544, USA}

\author{Zurab Guguchia}
\affiliation{Laboratory for Topological Quantum Matter and Advanced Spectroscopy(B7), Department of Physics, Princeton University, Princeton, NJ 08544, USA}
\affiliation{Laboratory for Muon Spin Spectroscopy, Paul Scherrer Institute, CH-5232 Villigen PSI, Switzerland}

\author{Songtian S. Zhang}
\affiliation{Laboratory for Topological Quantum Matter and Advanced Spectroscopy(B7), Department of Physics, Princeton University, Princeton, NJ 08544, USA}

\author{Tay-Rong Chang}
\affiliation{Department of Physics, National Cheng Kung University, Tainan 701, Taiwan}

\author{Hsin Lin}
\affiliation{Institute of Physics, Academia Sinica, Taipei 11529, Taiwan}

\author{Weiwei Xie}
\affiliation{Department of Chemistry, Louisiana State University, Baton Rouge, LA 70803, USA}

\author{M. Zahid Hasan}
\affiliation{Laboratory for Topological Quantum Matter and Advanced Spectroscopy(B7), Department of Physics, Princeton University, Princeton, NJ 08544, USA}
\affiliation{Princeton Institute for Science and Technology of Materials, Princeton University, Princeton, NJ 08544, USA}
\affiliation{Materials Science Division, Lawrence Berkeley National Laboratory, Berkeley, CA 94720, USA}

\author{Shuang Jia}
\email{gwljiashuang@pku.edu.cn}
\affiliation{International Center for Quantum Materials, School of Physics, Peking University, China}
\affiliation{Collaborative Innovation Center of Quantum Matter, Beijing 100871, China}
\affiliation{CAS Center for Excellence in Topological Quantum Computation, University of Chinese Academy of Sciences, Beijing 100190, China}
\affiliation{Beijing Academy of Quantum Information Sciences, West Building 3, No. 10 Xibeiwang East Road, Haidian District, Beijing 100193, China}

\date{\today}

\begin{abstract}

We study the anomalous Hall Effect (AHE) of single-crystalline $\mathrm{Co}_3\mathrm{Sn}_{2-x}\mathrm{In}_{x}\mathrm{S}_2$ over a large range of indium concentration $x$ from $0$ to $1$.
Their magnetization reduces progressively with increasing $x$ while their ground state evolves from a ferromagnetic Weyl semimetal into a nonmagnetic insulator.
Remarkably, after systematically scaling the AHE, we find that their intrinsic anomalous Hall conductivity (AHC) features an unexpected maximum at around $x=0.15$.
The change of the intrinsic AHC corresponds with the doping evolution of Berry curvature and the maximum arises from the magnetic topological nodal-ring gap.
Our experimental results show a larger AHC in a fundamental nodal-ring gap than that of Weyl nodes.

\end{abstract}
\maketitle
\section{INTRODUCTION}
The interplay between topology and magnetism is emerging as the new frontier in fundamental quantum physics \cite{nnano.2013.243,nature19820,Yan2017Topo,Hasan2017Discovery,2017physicsofquantummaterials,PhysRevB.98.224402,2019MagneticTI,2019JiaxinYin}.
A rich variety of topologically nontrivial states with time-reversal symmetry breaking have been realized in many material systems  \cite{PhysRevB.77.184402,PhysRevLett.102.186602,Chang167,RN23,RN22,Yin2018}.
One striking example is the magnetic Weyl semimetal (WSM) featuring the linearly dispersive band-touching points called Weyl nodes \cite{Yan2017Topo,Burkov2018Weyl,nmat4787,RN31}, analogous to the massless relativistic Weyl fermions in high energy physics.
As a signature of Wely nodes, large intrinsic anomalous Hall effect (AHE) has been observed in many candidates of magnetic WSMs \cite{PhysRevLett.88.207208, Burkov2018Weyl, RN20,RN13,RN2}.
Large AHE can stem from different types of hot zones of Berry curvature in momentum space such as Weyl nodes and topological nodal rings \cite{Belopolski1278}.
Understanding the relationship between the AHE and electronic structure in magnetic WSMs is crucial for designing functional quantum materials.
However the hot zones of Berry curvature are hard to be discerned by measuring the AHE because it is sensitively dependent on the Fermi level ($E_F$) and spin splitting.

Tuning the chemical potential by doping is a powerful tool to address the relationship between the AHE and Berry curvature change in a variety of magnetic materials \cite{Mathieu2004Scaling,Onose2006Doping,Takahashi2009Control,Lee2004Dissipationaless}.
For topological semimetals, however, this strategy is less proven efficient because other effects accompanied by the chemical doping such as electron scattering may also affect the AHE significantly.
In this paper we focus on a series of indium-substituted $\mathrm{Co}_3\mathrm{Sn}_2\mathrm{S}_2$  with a progressively changed chemical potential.
 $\mathrm{Co}_3\mathrm{Sn}_2\mathrm{S}_2$ is one of the very few compounds that has been confirmed to be magnetic WSM in experiment \cite{Morali1286,Liu1282}. This compound possesses $\mathrm{kagom\acute{e}}$ layers of Co atoms stacking in rhombohedral setting (Fig. 1(a)) and shows a half-metallic ferromagnetic (FM) ground state with a Curie temperature ($T_c$) of 177 K and spontaneous magnetic moment of 0.3 $\mu_B/\rm{Co}$. The electronic structure of $\mathrm{Co}_3\mathrm{Sn}_2\mathrm{S}_2$ forms two sets of linear band crossing points of nodal ring slightly above and below $E_F$ if spin-orbit coupling (SOC) is not considered. When SOC is taken into account, the Weyl nodes are  generated from the nodal rings split above $E_F$ while an indirect SOC gap forms below $E_F$ (Fig. 1(b)).  As a signature of the topological electron band, large intrinsic AHE has been observed in pristine $\mathrm{Co}_3\mathrm{Sn}_2\mathrm{S}_2$ \cite{RN13,RN2}.

We investigate the AHE in a series of $\mathrm{Co}_3\mathrm{Sn}_{2-x}\mathrm{In}_x\mathrm{S}_2$ where increasing indium substitution effectively lowers the chemical potential.
The indium substitution drives a transition from an FM WSM to a non-magnetic insulator in $\mathrm{Co}_3\mathrm{Sn}_{2-x}\mathrm{In}_{x}\mathrm{S}_2$, which can be well understood as the result of electron depopulating from the half-metallic band \cite{Interplay2015}. Remarkably, we find an enhanced anomalous Hall conductivity (AHC) up to 2500 $\mathrm{\Omega^{-1} cm^{-1}}$ when $x = 0.15$, compared to a monotonically damped magnetization by increasing $x$.~By scaling the AHC, we extract the intrinsic part which is well reproduced by the Berry curvature of $\mathrm{Co}_3\mathrm{Sn}_2\mathrm{S}_2$. Our results demonstrate that the SOC-induced nodal-ring gap can generate a larger AHC than that of the Weyl node in magnetic WSMs. This finding highlights a way to enhance the AHE in magnetic topological semimetals.


\begin{figure}
	\centering
	\includegraphics[scale=1]{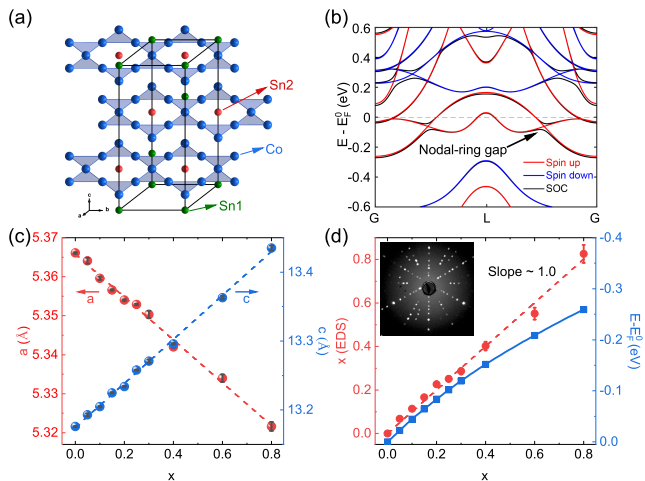}
	\caption{(a) Crystal structure of $\mathrm{Co}_3\mathrm{Sn}_2\mathrm{S}_2$ showing the $\mathrm{Co_3Sn}$ $\mathrm{kagom\acute{e}}$ layers. Sn1 and Sn2 represent the interlayer and intralayer sites of tin atoms, respectively. (b) Band structure of $\mathrm{Co}_3\mathrm{Sn}_2\mathrm{S}_2$ with/without SOC. (c) Variation of lattice parameters $a$ and $c$ with respect to $x$. (d) Measured indium concentration and inferred $E_F$ change with respect to $x$ (see also section IV of SI).~Inset:~Laue diffraction pattern.}
	\label{fig:mcmthesis-logo}
\end{figure}

\section{EXPERIMENTAL RESULTS}

Single crystals of $\mathrm{Co}_3\mathrm{Sn}_{2-x}\mathrm{In}_{x}\mathrm{S}_2$ were grown by the Bridgeman technique \cite{Holder2009Photoemission}. The stoichiometry and high quality of the crystals were confirmed by energy dispersive X-ray spectroscopy (EDS) and the structural refinement based on powder X-ray diffraction.
The lattice parameters $a$ and $c$ show linear dependence on the indium concentration for $0\leq x\leq 0.8$ (Fig.~\ref{fig:mcmthesis-logo}(c)), in good agreement with previously reported results \cite{RN6,Pielnhofer2014Half,Interplay2015,KASSEM2015208}.
Figure~\ref{fig:mcmthesis-logo}(d) shows that the $x$ determined by EDS equals the initial indium concentration in the molten solution, verifying the crystal's homogeneity.
An X-ray Laue back reflection photograph confirms the high quality and orientation of the single crystals (inset of Fig.~\ref{fig:mcmthesis-logo}(d)).~Our single crystal X-ray diffraction on representative samples confirms that the indium atoms preferentially substitute the interlayer tin atoms \cite{Zeitschrift2006,Pielnhofer2014Half,Interplay2015} and the $\mathrm{kagom\acute{e}}$ lattice remains intact when $x\leq 1$ \cite{Roth2013Effect}.

\begin{figure}
	\centering
	\includegraphics[scale=1]{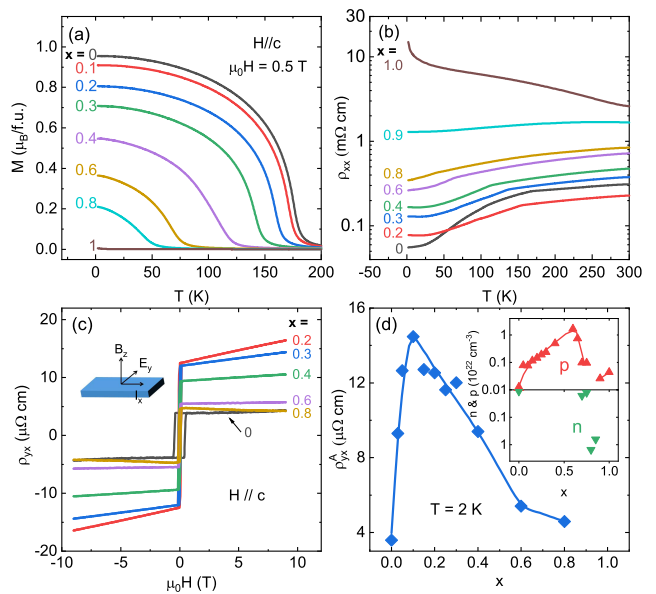}
	\caption{Temperature dependence of magnetization (a) and resistivity (b) for $\mathrm{Co}_3\mathrm{Sn}_{2-x}\mathrm{In}_{x}\mathrm{S}_2$. (c) Hall resistivity $\rho_{yx}$ as a function of magnetic field $\mu_0H$ at 2 K. (d)Anomalous Hall resistivity $\rho^A_{yx}$ at 2~K versus $x$. Inset: carrier density $n$ and $p$ at 2~K. The solid lines are guide to the eye.}
	\label{fig2}
\end{figure}

The magnetic and electrical transport properties for several single crystals with representative indium concentration are shown in  Fig.~\ref{fig2}.~While the $T_C=176 \pm 1~\rm{K}$ and the saturated magnetization at 2~K ($M_s=0.3~\mathrm{\mu_B/Co}$) for pristine $\mathrm{Co}_3\mathrm{Sn}_2\mathrm{S}_2$ are maximal for the whole series, both two values are monotonically suppressed by the indium substitution \cite{VAQUEIRO2009513,Weihrich2005CSS,Zeitschrift2006}.
When $x=1$, the sample exhibits non-magnetic behavior.
 In Fig.~\ref{fig2}(b), we plot temperature dependent resistivity for several $\mathrm{Co}_3\mathrm{Sn}_{2-x}\mathrm{In}_{x}\mathrm{S}_2$ samples.
 As previously reported \cite{RN2,RN13}, the resistivity at zero magnetic field ($\rho_{xx}$) shows a metallic profile with a kink around $T_C$ when $x\leq0.8$.
 Furthermore, the resistivity for the $x=1$ sample has an insulating profile, showing a metal-to-insulator transition with increasing x.
 This transition is the result of depopulation of the half-metallic band in which the indium substitution progressively lowers the chemical potential by removing one electron for each substituted tin atom (Fig.~\ref{fig:mcmthesis-logo}(d)) \cite{VAQUEIRO2009513,Weihrich2005CSS,Zeitschrift2006}.
The electronic structure of the series approximates to the rigid band of $\mathrm{Co}_3\mathrm{Sn}_2\mathrm{S}_2$, which gives us an opportunity for addressing the relation between the Berry curvature and AHE as the chemical potential changes in a wide range.

 The magnetic field dependence of  Hall resistivity $\rho_{yx}$ at 2~K for $\mathrm{Co}_3\mathrm{Sn}_{2-x}\mathrm{In}_{x}\mathrm{S}_2$ are shown in Fig.~\ref{fig2}(c).
 The observed $\rho_{yx}$ comprises a field dependent ordinary term and an anomalous term:
 \begin{equation}
  \rho_{yx} = \rho^O_{yx}(B)  + \rho^A_{yx} (M)
 \end{equation}
 where $B$ is the induction field and $\rho^A_{yx}$ is the anomalous Hall resistivity following the magnetic hysteresis loops with sharp switching.
 The $\rho^{O}_{yx}(B)$ for the pristine sample shows a nonlinear field dependence at 2~K, which is a typical two-band feature \cite{RN13}.
 For indium-substituted samples when $x < 0.7$, $\rho^O_{yx}(B) $ shows a linear dependence that can be interpreted by a single-band model.
 Figure \ref{fig2}(d) and inset plots $\rho^A_{yx}$ as the zero-field extrapolation of the high-field data and the extracted carrier densities, respectively.
The hole density ($p = 9.8 \times 10^{19}~\rm{cm}^{-3}$) is increased by a factor of $\sim$$170$ when $x$ increases from $0$ to $0.6$.
The Hall signals for $x = 0.7$ and $0.75$ show a two-band feature indicating a $p$-$n$ transition onset.
Further substitution significantly decreases the carrier density in insulating samples.

\begin{figure}
	\centering
	\includegraphics[scale = 1]{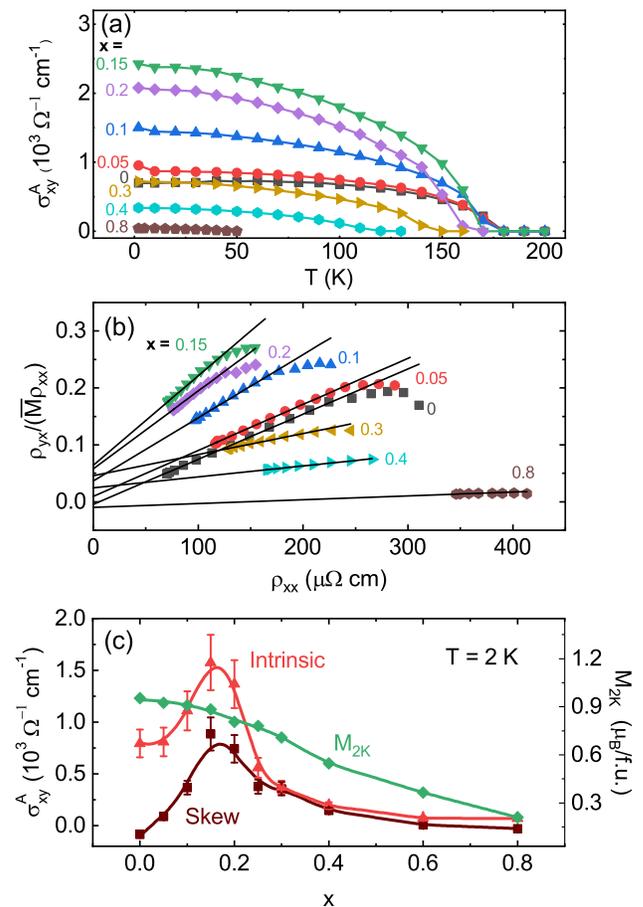}
	\caption{(a) Temperature dependent AHC for $\mathrm{Co}_3\mathrm{Sn}_{2-x}\mathrm{In}_{x}\mathrm{S}_2$. (b) $\rho^{A}_{yx}/(\overline{M}\rho_{xx})$ as a function of $\rho_{xx}$. In order to present all samples' data in one plot, we normalized the magnetization as $\overline{M} = M/M_{sat}$(2K). The solid lines across the data points are linear fit for low temperature regime.(c) Intrinsic and skew scattering AHC at 2~K for the whole series. Solid curves are guide to the eye. The error bar is estimated as $\pm$20\% in accordance to samples' dimension uncertainty.}
	\label{fig3}
\end{figure}

The anomalous Hall resistivity $\rho^A_{yx}$ shows a significant enhancement near $x=0.1$ and is then continuously suppressed until vanishing at $x>0.8$. The maximum of $\rho^A_{yx}$ is increased by a factor of 4 compared to pristine $\mathrm{Co}_3\mathrm{Sn}_2\mathrm{S}_2$.
It is clear that the variation of $\rho^A_{yx}$ at 2~K does not follow the trend of the magnetization and carrier density. The $p$-$n$ transition around $x=0.75$ also does not affect the sign of $\rho^A_{yx}$.
We calculate the AHC for the whole series at different temperatures by using $\sigma^A_{xy} = \frac{\rho^A_{yx}}{(\rho^A_{yx})^2+\rho_{xx}^2}$ (Fig.~\ref{fig3}(a)).
 The $\sigma^A_{xy}$ for pristine $\mathrm{Co}_3\mathrm{Sn}_2\mathrm{S}_2$ is $850~\pm~200~\mathrm{\Omega^{-1}cm^{-1}}$ at 2 K and remains nearly unchanged below 100 K, in good agreement with the previous results of the intrinsic AHE \cite{RN13}.~For $x=0.15$, $\sigma^A_{xy}$ becomes enhanced to 2500~$\pm$~500 $\mathrm{\Omega^{-1}cm^{-1}}$ at 2 K.
Unlike the pristine sample, $\sigma^A_{xy}$ for indium-substituted samples continuously increases with decreasing temperature, which indicates that part of $\sigma^A_{xy}$ is related with scattering events.

\section{DISCUSSION}

We now disentangle the intrinsic and extrinsic contributions to the AHE.
The anomalous Hall resistivity can be written as:
 \begin{equation}
  \rho^A_{yx} = (\alpha\rho_{xx}+\beta\rho^2_{xx})\cdot M(T)
 \end{equation}
in which the first term is linearly dependent on the resistivity, representing the extrinsic skew scattering contribution.
The second term is quadratically dependent on the resistivity and represents an intrinsic plus side-jump contribution in the AHE.
Both parts have a linear dependence on magnetization \cite{Zeng2006Linear,Nozires1973Asimple} and the parameters $\alpha$ and $\beta$ can therefore be obtained by plotting $\rho^A_{yx}/M\rho_{xx}$ versus $\rho_{xx}$, as shown in Fig.~\ref{fig3}(b).
The data shows good linearity at low temperatures for all samples and we can extract the intercept ($\alpha$) and slope ($\beta$) and calculate the skew scattering and intrinsic plus side-jump AHC at 2~K.
Figure \ref{fig3}(c) shows that the intrinsic plus side-jump AHC has a maximum up to 1500 $\pm$ 300 $\Omega^{-1}\mathrm{cm}^{-1}$ at $x=0.15$, approximately twice as large as the pristine case.
As comparison, the skew-scattering AHC ($\sigma_{xy}^{sk}$) also shows a broad maximum near $x=0.15$ and then slowly decreases with increasing $x$.
Previous studies have reported the AHE in a variety of chemical compositions \cite{Mathieu2004Scaling,Onose2006Doping,Takahashi2009Control,Lee2004Dissipationaless}, but the non-monotonic change of the AHC is less observed in topological semimetals.

\begin{figure}
	\centering
	\includegraphics[scale=1]{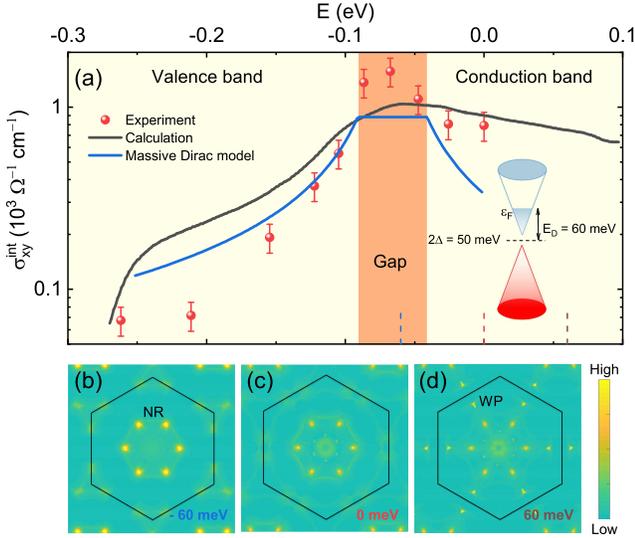}
	\caption{(a) Black solid line: calculated  $\sigma^{int}_{xy}$ for $\mathrm{Co}_3\mathrm{Sn}_2\mathrm{S}_2$ with respect to the chemical potential. Red dots with error bar: experimental results of the intrinsic plus side-jump AHC for $\mathrm{Co}_3\mathrm{Sn}_{2-x}\mathrm{In}_{x}\mathrm{S}_2$ at 2~K. The chemical potential is inferred from rigid band approximation (see Fig. \ref{fig:mcmthesis-logo}(d)). Blue solid line: the AHC for massive Dirac model. (b), (c) and (d): Berry curvature distribution projected to the $k_x$-$k_y$ plane for  $\mathrm{Co}_3\mathrm{Sn}_2\mathrm{S}_2$ at the energy of -60, 0 and 60 meV, respectively. The color bar on right presents the Berry curvature intensity. NR and WP represent nodal ring and Weyl point, respectively.}
	\label{fig4}
\end{figure}
To shed light on the non-monotonic change in the AHC, we compare the extracted intrinsic plus side-jump AHC with the calculated energy-dependent AHC ($\sigma^{int}_{xy}$) in accordance to the Berry curvature of pristine  $\mathrm{Co}_3\mathrm{Sn}_2\mathrm{S}_2$ in Fig.~\ref{fig4}(a).
The $\sigma^{int}_{xy}$ has a similar profile with respect to energy as the experimental data: it is enhanced below $E_F$ with a maximum of $\sim$$1000$ $\mathrm{\Omega^{-1} cm^{-1}}$ and then drops to nearly zero at $-270$ meV.
The large $\sigma^{int}_{xy}$ is attributed to hot zones of Berry curvature in the Brillouin zone.~Figure \ref{fig4}(b)-(d) show the Berry curvature distribution projected to the $k_x$-$k_y$ plane at three different energies of -60, 0 and 60 meV, respectively.
These energy levels correspond to the $E_F$ of the SOC gap at high-symmetry line, for the pristine sample and that of the Weyl nodes respectively.
The color maps show different origins of hot zones: the main contributions from the Weyl nodes and the gapped nodal rings, plus some negligible contribution from other bands.~The contributions from the Weyl nodes and nodal rings are comparable as the $E_F$ is located at 60 meV. With decreasing $E_F$, the Berry curvature contribution from the nodal rings is enhanced, compensating the reduced contribution from the Weyl nodes.
The total $\sigma^{int}_{xy}$ has a maximum around -60 meV, in agreement with our transport experiment.

Our measurements and calculations show that the small SOC gap in  $\mathrm{Co}_3\mathrm{Sn}_2\mathrm{S}_2$ generates a large AHC which is comparable to a quantized value.
To illustrate this point clearly, we compare this result to the AHC in a massive Dirac model \cite{Sinitsyn2007Anomalous} which can be realized in a $\mathrm{kagom\acute{e}}$ lattice of the magnetic atoms \cite{Xu2015Intrinsic}.
The corresponding AHC equals $\sigma_{xy}^0 = 3\cdot\frac{e^2}{h\cdot c} = 881~\Omega^{-1} \rm{cm}^{-1}$ with the quantized Hall conductance $e^2/h = 3.87 \times 10^{-5}~\Omega^{-1}$ and $c = 13.176~\rm{\AA}$ for a fictitious lattice hosting massive Dirac cone. The factor 3 is introduced because the unit cell of  $\mathrm{Co}_3\mathrm{Sn}_2\mathrm{S}_2$ contains three layers of Co $\mathrm{kagom\acute{e}}$ lattice.
This value is slightly smaller than $\sigma^{int}_{xy}$ and the experimental results for $\mathrm{Co}_3\mathrm{Sn}_{2-x}\mathrm{In}_{x}\mathrm{S}_2$ in the region of $-60\pm 25$ meV.
In a large energy scale, however, the profile of AHC in this simple massive Dirac model is close to the experimental result.

Part of the discrepancy between the calculated $\sigma^{int}_{xy}$ and experimental results may be attributed to the side-jump contribution which also has a quadratic dependence of $\rho_{xx}$ and is therefore difficult to distinguish from the intrinsic one.
However the side-jump contribution should be smaller than $e^2/hc$ and the intrinsic contribution is dominant in the total AHC \cite{Onoda2006Intrinsic}. The side-jump contribution can be estimated as $\frac{e^2}{ha}\frac{\varepsilon_{SO}}{E_F}$\cite{Nozi1973_Asimple} where $\varepsilon_{SO}$ is the strength of the SOC. Given $\varepsilon_{SO}$ is about 50 meV, we estimate $\sigma_{xy}^{sj}$ about 50 $\rm{\Omega^{-1}cm^{-1}}$, much less than the intrinsic contribution.
Another point of concern is the $\sigma_{xy}^{sk}$ which also exhibits a broad maximum near $x = 0.15$.
Due to multi-variable complexity including spin-orbit interaction, scattering strength and disorder density, calculating the energy dependence of $\sigma_{xy}^{sk}$ in $\mathrm{Co}_3\mathrm{Sn}_{2-x}\mathrm{In}_{x}\mathrm{S}_2$ is difficult.
We briefly discuss the origin of non-monotonic change of $\sigma_{xy}^{sk}$ as below.
As shown in Fig.~S5 in SI, $\sigma_{xy}^{sk}$ is linearly dependent on the carrier mobility $\mu$ when $x \geq 0.15$.
In contrast $\sigma_{xy}^{sk}$ starkly deviates from the linear dependence on the mobility as long as $x < 0.15$.
This deviation is not unexpected because the chemical potential crosses the SOC gap when $x$ is about $0.15$.
Therefore the non-monotonic change of  $\sigma_{xy}^{sk}$ in $\mathrm{Co}_3\mathrm{Sn}_{2-x}\mathrm{In}_{x}\mathrm{S}_2$ reflects the fact that Fermi energy  across the SOC gap near $x=0.15$.
We notice that the discontinuous change in $\sigma_{xy}^{sk}$ near the SOC gap was shown for some model systems \cite{Onoda2008_QuantumTransport}.~The detailed electronic structure for this alloy system asks for elaborations in the future.

\section{CONCLUSION}

In summary, we have investigated the AHE in a series of $\mathrm{Co}_3\mathrm{Sn}_{2-x}\mathrm{In}_{x}\mathrm{S}_2$ single crystals which change from an FM WSM to a non-magnetic insulator as $x$ increases.
This transition can be well described as the result of band depopulation by removing one electron per formula unit from the pristine $\mathrm{Co}_3\mathrm{Sn}_2\mathrm{S}_2$.
We separate the intrinsic and skew-scattering AHC and find that both do not follow the change of magnetization or carrier density.
The intrinsic AHC obtained in experiment is consistent with the non-monotonic change of Berry curvature of $\mathrm{Co}_3\mathrm{Sn}_2\mathrm{S}_2$ dependent on $E_F$.~The SOC-induced nodal-ring gap contributes a strong hot zone of Berry curvature which induces an enhanced AHC. Our work clarifies the source of the large AHC in the first experimentally confirmed magnetic WSM $\mathrm{Co}_3\mathrm{Sn}_2\mathrm{S}_2$ and provides a path to enhance the AHE in other topological materials.

\section*{ACKNOWLEDGMENTS}

The authors thank Qian Niu, Zhe Hou and Jia Li for helpful discussions, Wenlong Ma and Yanfang Li for assistance in modifications of figures. Shuang Jia was supported by the National Natural Science Foundation of China No.~11774007, No.~U1832214, the National Key R$\&$D Program of China (2018YFA0305601) and the strategic Priority Research Program of Chinese Academy of Sciences, Grant No. XDB28000000. The work at LSU is supported by Beckman Young Investigator award. T.-R.C. was supported from Young Scholar Fellowship Program by Ministry of Science and Technology (MOST) in Taiwan, under MOST Grant for the Columbus Program MOST108-2636-M-006-002, National Cheng Kung University, Taiwan, and National Center for Theoretical Sciences (NCTS), Taiwan.

\bibliographystyle{plain}

\end{document}